\documentclass[aps,prl,reprint,superscriptaddress,twocolumn,showkeys,amsmath,amssymb,longbibliography]{revtex4-1}
\usepackage[english]{babel}
\usepackage{amsmath,amssymb,bbm,mathrsfs,bm,braket,color,graphicx,comment,amsfonts,dsfont}
\usepackage[colorlinks,citecolor=blue,urlcolor=blue]{hyperref}
\usepackage[mathscr]{euscript}
\usepackage[normalem]{ulem}
\usepackage{comment}

\begin{document}
\title{Anomalous levitation and annihilation in Floquet topological insulators}
\author{Hui Liu}
\affiliation{IFW Dresden and W{\"u}rzburg-Dresden Cluster of Excellence ct.qmat, Helmholtzstrasse 20, 01069 Dresden, Germany}
\author{Ion Cosma Fulga}
\affiliation{IFW Dresden and W{\"u}rzburg-Dresden Cluster of Excellence ct.qmat, Helmholtzstrasse 20, 01069 Dresden, Germany}
\author{J\'{a}nos K. Asb\'{o}th}
\affiliation{Institute for Solid State Physics and Optics, Wigner Research Centre for Physics, H-1525 Budapest P.O. Box 49, Hungary}
\affiliation{Department of Theoretical Physics, Budapest University of Technology and Economics}

\begin{abstract}
Anderson localization in two-dimensional topological insulators takes place via the so-called \emph{levitation and pair annihilation} process.
As disorder is increased, extended bulk states carrying opposite topological invariants move towards each other in energy, reducing the size of the topological gap, eventually meeting and localizing.
This results in a topologically trivial Anderson insulator.
Here, we introduce the anomalous levitation and pair annihilation, a process unique to periodically-driven, or Floquet systems.
Due to the periodicity of the quasienergy spectrum, we find it is possible for the topological gap to \emph{increase} as a function of disorder strength.
Thus, after all bulk states have localized, the system remains topologically nontrivial, forming an anomalous Floquet--Anderson insulator (AFAI) phase.
We show a concrete example for this process, adding disorder via onsite potential ``kicks'' to a Chern insulator model.
By changing the period between kicks, we can tune which type of (conventional or anomalous) levitation-and-annihilation occurs in the system.
We expect our results to be applicable to generic Floquet topological systems and to provide an accessible way to realize AFAIs experimentally, without the need for multi-step driving schemes.
\end{abstract}
\maketitle
\emph{Introduction} ---
In fully coherent systems, disorder leads to a loss of metallic conduction and a transition to a localized state: the Anderson insulator (AI) \cite{Anderson1958, Evers2008}.
In three dimensions, this is a gradual process.
For small disorder, each energy band is split by so-called mobility edges to a middle part with extended states and outer parts with localized states, whose localization length diverges at the mobility edges.
As disorder is increased, the central, extended part of each band shrinks, and eventually the two mobility edges meet and the bands become localized.
In contrast, in generic one- and two-dimensional systems, already an arbitrarily weak disorder is enough to localize all bulk states.

Shortly after the discovery of the quantum Hall effect \cite{Klitzing}, it was realized that two-dimensional Chern insulators also require a finite amount of disorder to localize, but through a different type of transition \cite{Laughlin1984}.
In bands with a nonzero Chern number, although almost all bulk eigenstates can be (and are) exponentially localized, the localization length diverges at isolated energies: extended states ``carry the Chern number'' \cite{Halperin1982aa, Thouless1984, Thonhauser2006}.
As found by Laughlin \cite{Laughlin1984}, the extended bulk states carrying opposite Chern numbers ``levitate'' towards each other in energy when disorder is gradually increased, and eventually ``annihilate'' pairwise, so Anderson localization sets in \cite{Prodan2011}.
Since the topological edge states only occur within the mobility gap between the extended bulk states, the pair annihilation leads to a topologically trivial system.

\begin{figure}
\centering
\includegraphics[width=\columnwidth]{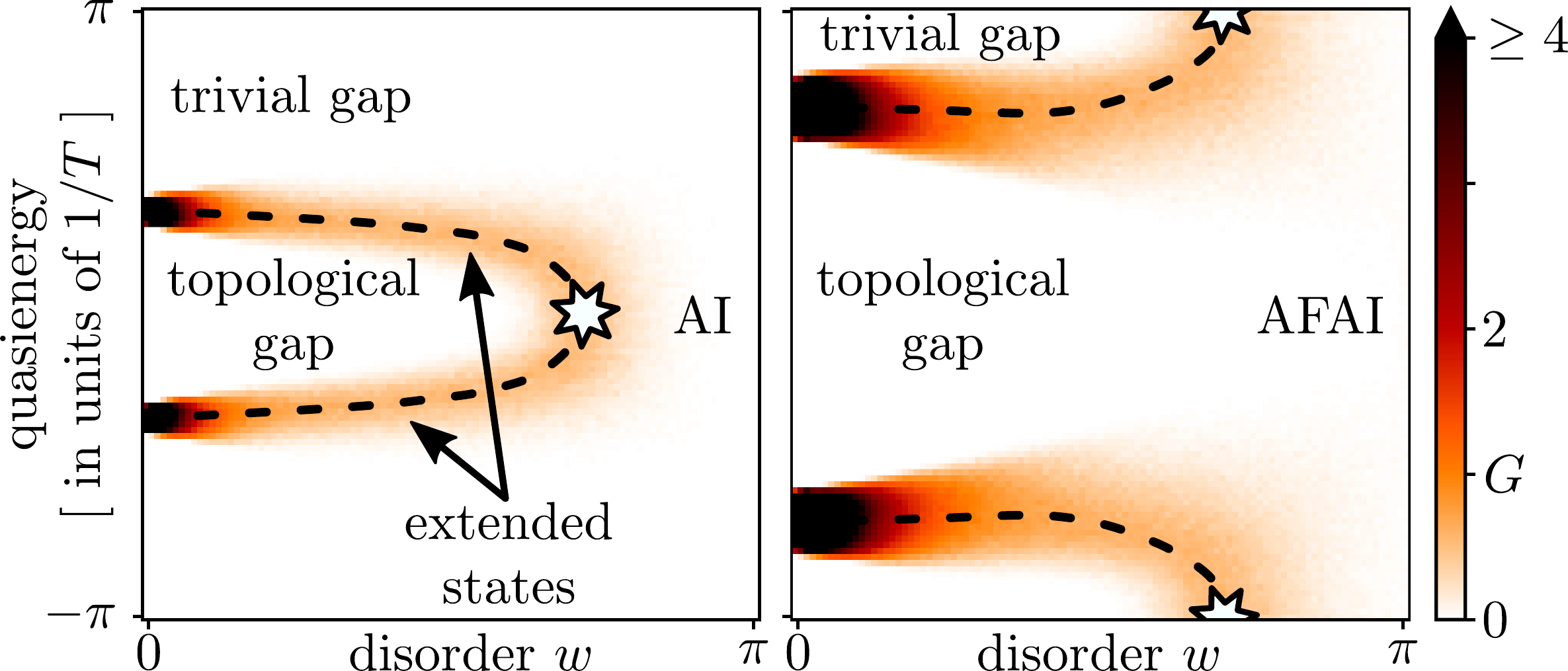}
\caption{
When increasing disorder, the extended states separating topological and trivial quasienergy gaps levitate towards each other, eventually annihilating pairwise (white star).
The conventional form of this process leads to the elimination of the topological gap, resulting in a trivial Anderson insulator (AI, left panel).
The anomalous levitation and pair annihilation, in which the trivial gap shrinks, leaves behind an anomalous Floquet--Anderson insulator (AFAI, right panel).
The background colors are numerically obtained by computing the transmission, as explained in the main text \cite{params1}.
\label{fig:sketch_anomalous_levitation}
}
\end{figure}

In the last decade, it was found that robust extended edge states and complete bulk Anderson localization can coexist in periodically driven systems, so-called anomalous Floquet--Anderson insulators (AFAI) \cite{Titum2016, Kundu2020, Nathan2017}.
Even if all quasienergy bands have zero Chern numbers, Floquet insulators can have topologically protected chiral edge states, which wind in quasienergy \cite{Kitagawa2010, Rudner2013, Nathan2015,
Else2016, Nathan2017, Mukherjee2017, Maczewsky2017, Quelle2017,
Fulga2019, Wintersperger2020}.
Since the bands are trivial, arbitrarily weak disorder leads to a fully localized bulk, while leaving the chiral edge states extended \cite{Titum2016, Kundu2020, Nathan2017}: 
There is no levitation-and-annihilation in such anomalous Floquet topological insulators.

In this paper, we revisit Laughlin's result on Anderson localization in the context of Floquet Chern insulators \cite{Oka2009, Inoue2010, Lindner2011, Gu2011, Kitagawa2011, Katan2013, Iadecola2013, Cayssol2013, Goldman2014, Grushin2014, Kundu2014, Titum2015, Bukov2015}.
The quasienergy bands of these systems carry Chern numbers, and hence we expect a levitation-and-annihilation scenario \cite{Roy2016}.
However, even in the simplest, two-band models, there are two different ways in which extended states carrying opposite Chern numbers can meet and annihilate.
Due to the periodic spectrum, the extended states can levitate towards each other by reducing the size of the topological gap (the conventional scenario) or by increasing it instead (see Fig.~\ref{fig:sketch_anomalous_levitation}).
Thus, disorder can induce a transition from a Floquet Chern insulator not only to an AI, but also to an AFAI.
We show this in the following using a toy model for a Floquet Chern insulator, in which both scenarios of levitation and annihilation happen, and find a simple rule of thumb for when to expect either scenario.

\begin{figure}
\centering
\includegraphics[width=1\linewidth]{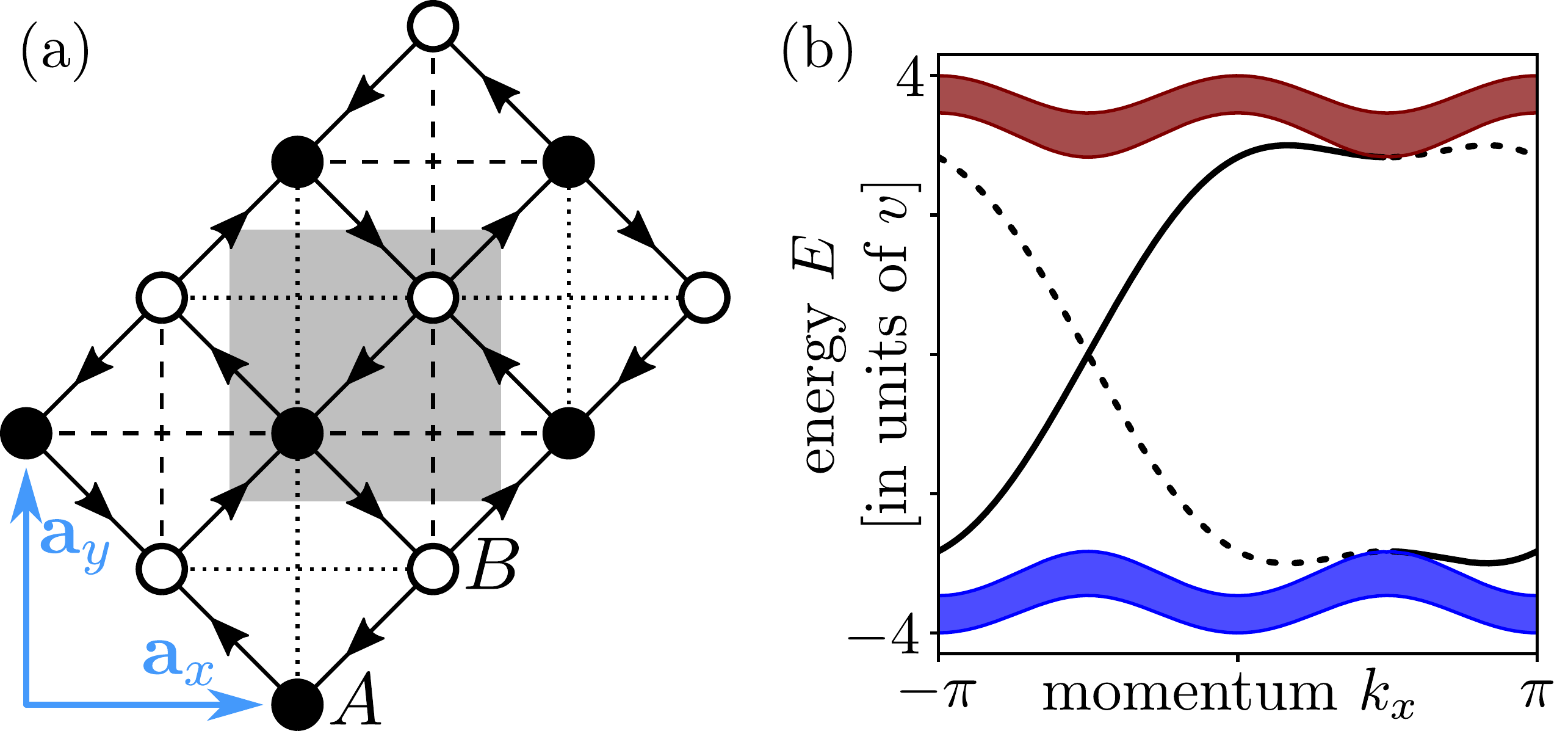}
\caption{
(a) The model Hamiltonian:
One unit cell (gray square) contains two sites, belonging to the $A$ and $B$ sublattice (filled/empty circle).
Bravais vectors ${\bf a}_{x,y}$ are indicated as blue arrows.
Nearest-neighbor hopping amplitudes along/against the arrows are $v_1 (1+i)$ and $v_1 (1-i)$, respectively.
Next nearest neighbor hopping amplitudes are $+v$ along the dashed lines and $-v$ along the dotted lines.
(b) Dispersion relation computed for $v_1=v$ in a ribbon geometry, infinite along ${\bf a}_x$ and consisting of $20$ unit cells in the ${\bf a}_y$ direction.
Top/bottom edge modes are shown as dashed/solid black lines, respectively.
}
\label{fig:static}
\end{figure}

\emph{System} --- 
We consider a tight-binding model on a square lattice, with Hamiltonian \cite{Neupert2011}
\begin{align}
  \label{eq:ham}
  \hat{H}_0 =& v_1 (1+i) \sum_\mathbf{r} \Big( \ket{A_{{\bf r}}}\bra{B_{{\bf r}}} + 
        \ket{A_{{\bf r}+{\bf a}_x+{\bf a}_y}}\bra{B_{{\bf r}}} + \nonumber\\
 & \hspace{45pt} \ket{B_{{\bf r}}}\bra{A_{{\bf r}+{\bf a}_x}} 
          + \ket{B_{{\bf r}}}\bra{A_{{\bf r}+{\bf a}_y}} \Big)  \nonumber\\
 & + v \sum_\mathbf{r} \Big( \ket{A_{{\bf r}}}\bra{A_{{\bf r}+{\bf a}_x}}
             - \ket{A_{{\bf r}}}\bra{A_{{\bf r}+{\bf a}_y}} + \nonumber\\
 & \hspace{30pt} \ket{B_{{\bf r}}}\bra{B_{{\bf r}+{\bf a}_y}}
            - \ket{B_{{\bf r}}}\bra{B_{{\bf r}-{\bf a}_x}} \Big) + {\rm h.c},
\end{align}
where $\ket{A_\mathbf{r}}$ and $\ket{B_\mathbf{r}}$ denote sites on the $A$ and $B$ sublattice in the unit cell (see Fig.~\ref{fig:static}) with coordinates $\mathbf{r}= N_x \mathbf{a}_x + N_y \mathbf{a}_y$, with $N_x,N_y\in \mathbb{Z}$.
We measure energy in units of $v$, time in units of $1/v$ ($\hbar=1$ throughout), and distance along $x$ and $y$ in units of $\left|{a}_x\right|$ and
  $\left|{a}_y\right|$.
The two energy bands are symmetric around $E=0$ because $\mathrm{tr}\hat{H}_0(k)=0$.
For the most part of this paper we will use $v_1=v$, where the bands have Chern numbers $\pm 1$, and thus the gap separating them is topological, \emph{i.e.}, hosts one branch of chiral edge states.
Here, as Fig.~\ref{fig:static}(b) shows, the bands are relatively flat: their bandwidths, $1.17 v$ are much smaller than the band gap $\Delta=5.66 v$.

We add disorder to the hopping model in the form of periodic onsite potential kicks,
\begin{align}
  \label{eq:H_t_def}
  \hat{H}(t) &= \hat{H}_0
  + w \hat{H}_\mathrm{dis} \sum_{n\in\mathbb{Z}}\delta(t-nT),
\end{align}
The time period $T$ separates the kicks, which have a strength $w\in\mathbb{R}$ and are spatially random,
\begin{align}
  \hat{H}_\mathrm{dis} &= \sum_{\pmb{r}}\left(
  \xi_{\mathbf{r},A} \ket{A_\mathbf{r}} \bra{A_\mathbf{r}}
  + \xi_{\mathbf{r},B} \ket{B_\mathbf{r}}\bra{B_\mathbf{r}} \right),
\end{align}
with $\xi_{\pmb{r},A/B}$ random numbers drawn independently for each lattice site, uniformly distributed with $-1\le \xi \le 1$. Note that the delta function in Eq.~\eqref{eq:H_t_def} has units of inverse time, or energy, such that both $w$ and $\xi$ are dimensionless.

\begin{figure}
\centering
\includegraphics[width=0.8\columnwidth]{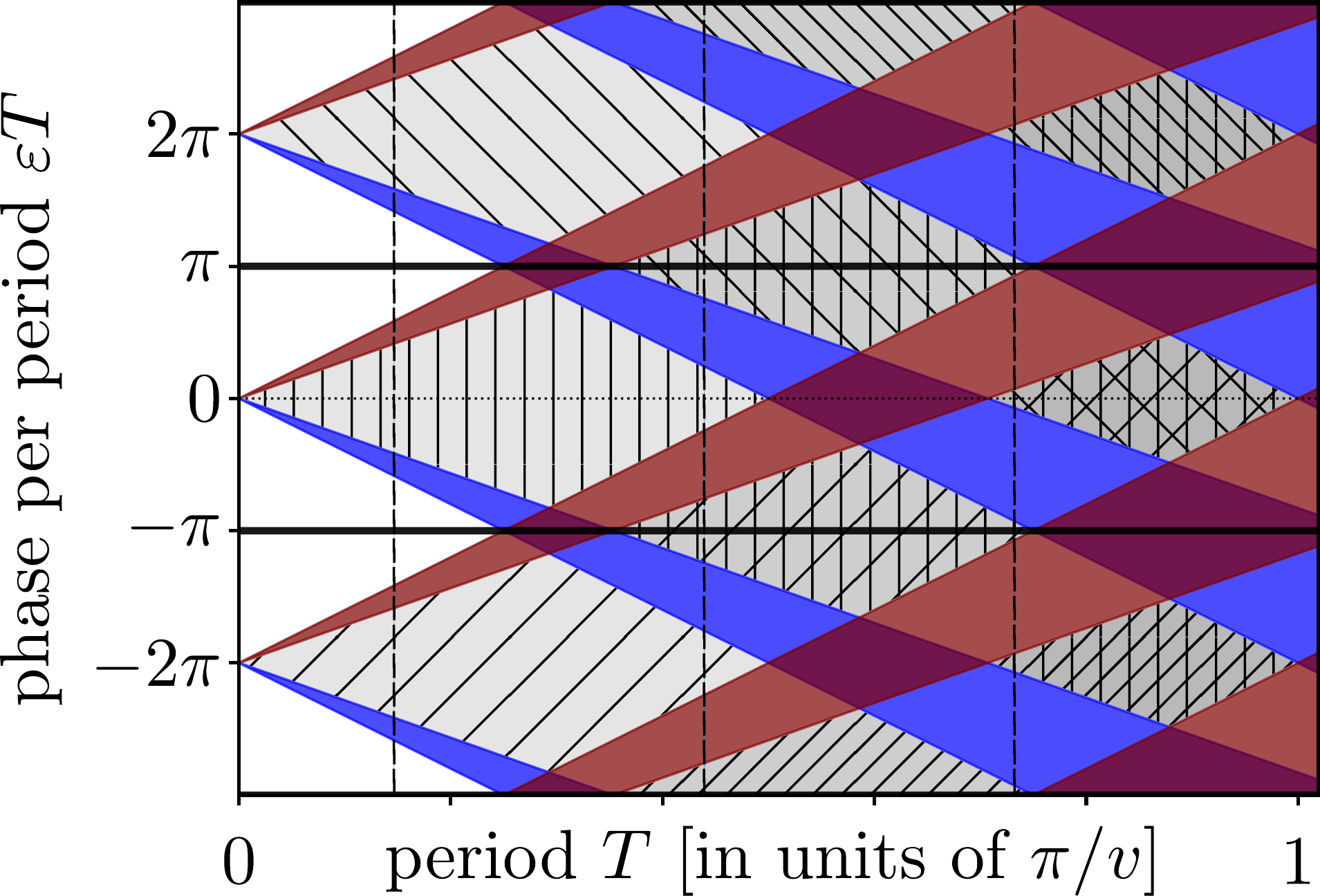}
\caption{ Floquet spectrum of Eq.~\eqref{eq:floquet_operator} with
  $v_1=v$ in the clean limit, $w=0$, in a repeated Floquet zone
  scheme. Tuning the time period $T$ has no physical effects here, it
  only changes our description of the same physics, described by the
  constant Hamiltonian $\hat{H}_0$ -- as in the first
  step of the nearly free electron approximation of crystalline
  solids.  Hatching indicates the presence of edge states in a gap.
  Increasing the period $T$, bands cross over between Floquet zones,
  delimited by thick horizontal lines.  This results in a sequence of
  topological phase transitions: gaps are closed and reopened (even if
  in this undriven case no transitions between bands happen), with the
  number of edge states in them increasing by 2 each time
  \cite{suppmat}.  Vertical dashed lines indicate periods for which
  the gaps are equal, $T=nT_c$ with $n=1,3,5$.
\label{fig:floquet_spectrum}
}
\end{figure}

The Floquet spectrum is the spectrum of the Floquet operator, $\hat{\mathcal{F}}$, the time evolution operator of one period,
\begin{align}\label{eq:floquet_operator}
  \hat{\mathcal{F}}&= \mathcal{T} e^{-i\int_0^T \hat{H}(t')dt'}
  = e^{-i w \hat{H}_\mathrm{dis}} e^{-i\hat{H}_0 T},
\end{align}
where $\mathcal{T}$ denotes time ordering.
Eigenstates of $\hat{\mathcal{F}}$ -- the Floquet eigenstates -- pick up phase factors of $e^{-i\varepsilon T}$ during each period, where the quasienergy $\varepsilon$ takes values in $[ -\pi/T,\pi/T ] $ -- the Floquet zone, in analogy with the Brillouin zone.
In the limit of maximal disorder, $w=\pi$, the kicks randomize quasienergy completely, meaning that all disorder-averaged properties of the model are independent of quasienergy.

As a first step to understanding the effects of the periodic kicks, we
take the limit of vanishing kick strength $w\to 0$, similarly to the
way lattice effects are treated in the nearly free electron model of
crystalline solids. This amounts to time evolution
using the static Hamiltonian, $\hat{H}_0$, but calculating the effects
only at integer multiples of a time period $T$. In the absence of
kicks, the time period $T$ does not change any of the physical
properties of the system, only the type of information we can extract
from it: any eigenstate of $\hat{H}_0$, with energy $E$, is also an
eigenstate of $\hat{\mathcal{F}}$. The corresponding quasienergy is
$\varepsilon=E$, projected into the first Floquet zone, \emph{i.e.},
$\varepsilon = \left[ (ET + \pi) \mod (2\pi) - \pi \right]/T$. Like in
the nearly free electron model, we will use a repeated Floquet zone
description here, and for simplicity sometimes argue using the ``phase
per period'' $\varepsilon T$ -- which is the same as the quasienergy
$\varepsilon$ measured in units of $1/T$.

\emph{Topology} --- Even in the limit of vanishing kick strength, $w\to 0$, the time period $T$ can be used to tune the topological invariants of the system, the winding numbers $\mathcal{W}$ \cite{Rudner2013} of the quasienergy gaps.
To see this, we follow the quasienergy bands in a repeated Floquet zone scheme, in Fig.~\ref{fig:floquet_spectrum}.
For $T<0.25\pi/v$, all the energy spectrum of $\hat{H}_0$ fits in the first Floquet zone, including edge states.
Thus the quasienergy spectrum consists of Floquet replicas of the lower and upper band, together with the edge states between them, and the gap around $\varepsilon = 0$ is topological, whereas the gap at the Floquet zone boundary, $\varepsilon = \pi/T$, is trivial.
As $T$ is increased, the gap at $\varepsilon= 0$ grows relative to the gap at $\varepsilon=\pi/T$, and eventually overtakes it at a critical period time,
\begin{align}
  \label{eq:T_c}
  T_c &= \frac{\pi}{2 E_{1/2}} = \frac{\pi}{(4+2\sqrt{2})v} \approx \frac{0.15 \pi}{v},
\end{align}
where $E_{1/2}$ is the band center, the average of the minimum and maximum energies of the upper band \cite{suppmat}.
At around $T \approx 2 T_c$ the bands cross the Floquet zone boundaries: the $\varepsilon = \pi/T$ gap closes, and when it reopens, hosts edge states coming from the first as well as from the second Floquet zones.
Thus, the winding number of this gap changes from 0 to 2. 
Further increasing $T$ leads to a sequence of similar transitions at $T=2nT_c$, at $\varepsilon=0$ for even $n$ and $\varepsilon=\pi/T$ for odd $n$.

\emph{Disorder} --- 
To investigate how disorder $w$ affects the system we calculate the two-terminal transmission $G$ using the Kwant code \cite{Groth2014, suppmat}.
We consider a finite system of $L\times L$ unit cells, with either periodic or open boundary conditions in the ${\bf a}_x$ direction, and semi-infinite leads attached at the top and bottom, modeled as absorbing terminals at $N_y=1$ and $N_y=L$.
The two-terminal scattering matrix $\hat{S}$ reads \cite{Fulga2016}
\begin{eqnarray}
\hat{S}(\varepsilon) =
\hat{P} [1-e^{i\varepsilon}\hat{\mathcal{F}} (1-\hat{P}^{T} \hat{P})]^{-1}
e^{i\varepsilon} \hat{\mathcal{F}}
\hat{P}^{T}
\label{eq:fsm},
\end{eqnarray}
where $\hat{P}$ is the projection operator onto the absorbing terminals.
The total transmission $G(\varepsilon)$ can be extracted from the scattering matrix $S(\varepsilon)$:
\begin{align}
  \label{eq:smatrix}
G&={\rm
  tr}(\mathfrak{t}^\dag\mathfrak{t}),&
S&=\begin{pmatrix}
        \mathfrak{r} & \mathfrak{t}' \\
        \mathfrak{t} & \mathfrak{r}'
\end{pmatrix},
\end{align}
where $\mathfrak{r}^{(\prime)}$ and $\mathfrak{t}^{(\prime)}$ are the blocks containing probability amplitudes for back-reflection, or transmission between the two terminals, respectively, whose dependence on $\varepsilon$ was suppressed for readability.
With periodic boundary conditions and at maximal disorder, $w=\pi$, the transmission is quasienergy-independent, such that a vanishing transmission at any value of $\varepsilon$ indicates total localization of all bulk states.
Changing to open boundary conditions, for an AFAI phase with topological invariant $\mathcal{W}$, topologically protected edge states will appear, constitute completely open channels for transport, and contribute integer values to the total transmission, with $G(\varepsilon) = \left|{\mathcal{W}}\right|$ for all $\varepsilon$. 
Alternatively, the invariant $\mathcal{W}$ can be obtained as the winding of the determinant of the reflection part $\mathfrak{r}$ of the scattering matrix \cite{fulga2012scattering, Fulga2016, suppmat}.

\begin{figure}
\centering
\includegraphics[width=1\linewidth]{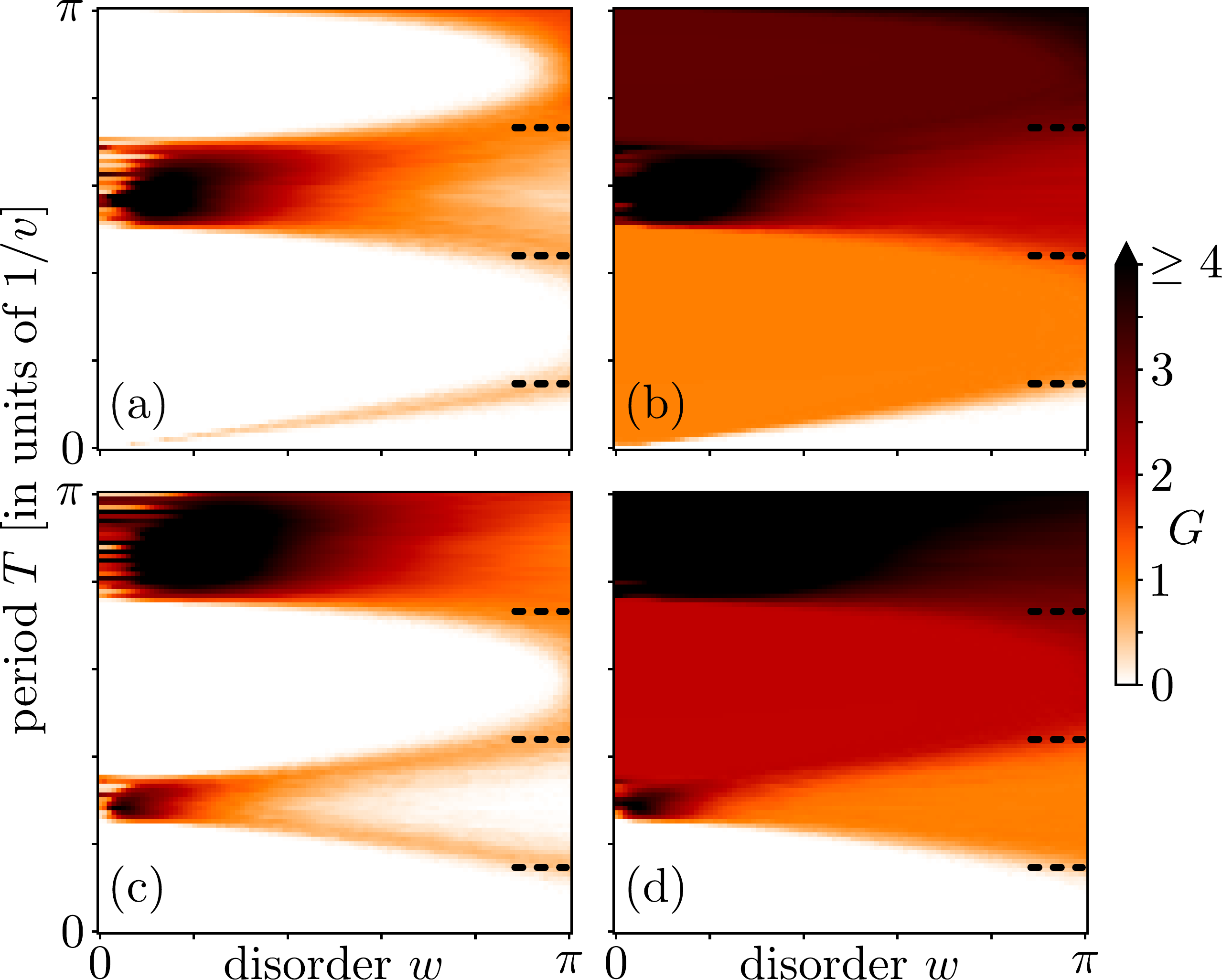}
\caption{
Phase diagram of the transmission $G$ depending on disorder $w$ and period $T$, at quasienergy $\varepsilon=0$ (a-b) and $\varepsilon=\pi$ (c-d). 
The boundary conditions along $x$ are periodic in (a) and (c), and open in (b) and (d). 
Transmissions are calculated for a system size of $20\times 20 $, with $v_1=v$, and by averaging over $50$ disorder realizations. 
Dashed lines show the analytically predicted phase transition points, as described in the main text.
}
\label{fig:tw}
\end{figure}

The calculated values of the transmission (see Fig.~\ref{fig:tw}) show that, depending on the period $T$, increasing disorder strength can lead to a transition to an AI or an AFAI, with either the usual or the anomalous levitation-and-annihilation scenarios of Fig.~\ref{fig:sketch_anomalous_levitation}.
For $T\ll 0.25\pi/v$, we find the first scenario: in Fig.~\ref{fig:tw}(a) the mobility gap at $\varepsilon=0$ closes and reopens as two extended states carrying the Chern numbers meet and annihilate, while the winding number of this gap changes from 1 to 0, as evidenced in Fig.~\ref{fig:tw}(b).
We show an example in more detail, for $T=0.1\pi/v$, in Fig.~\ref{fig:sketch_anomalous_levitation}.
For a range of period times ($0.15 \pi/v < T < 0.45 \pi/v$), we find the second scenario, a transition to AFAI via an anomalous levitation and annihilation: in Fig.~\ref{fig:tw}(c), the mobility gap closes and reopens at $\varepsilon=\pi/T$, and edge transmission indicates a winding number of 1 in the $w=\pi$ limit, in Fig.~\ref{fig:tw}(b) and (d).
An example, with $T=0.2\pi/v$, is shown in Fig.~\ref{fig:sketch_anomalous_levitation}.
For longer drive periods we observe hints of transitions to AFAI phases with higher winding numbers: around $T\approx 0.5\pi/v$, of $\mathcal{W}=2$, and around $T=0.9\pi/v$, possibly $\mathcal{W}=3$, although with substantial finite size effects.

\begin{figure}
\centering
\includegraphics[width=1\linewidth]{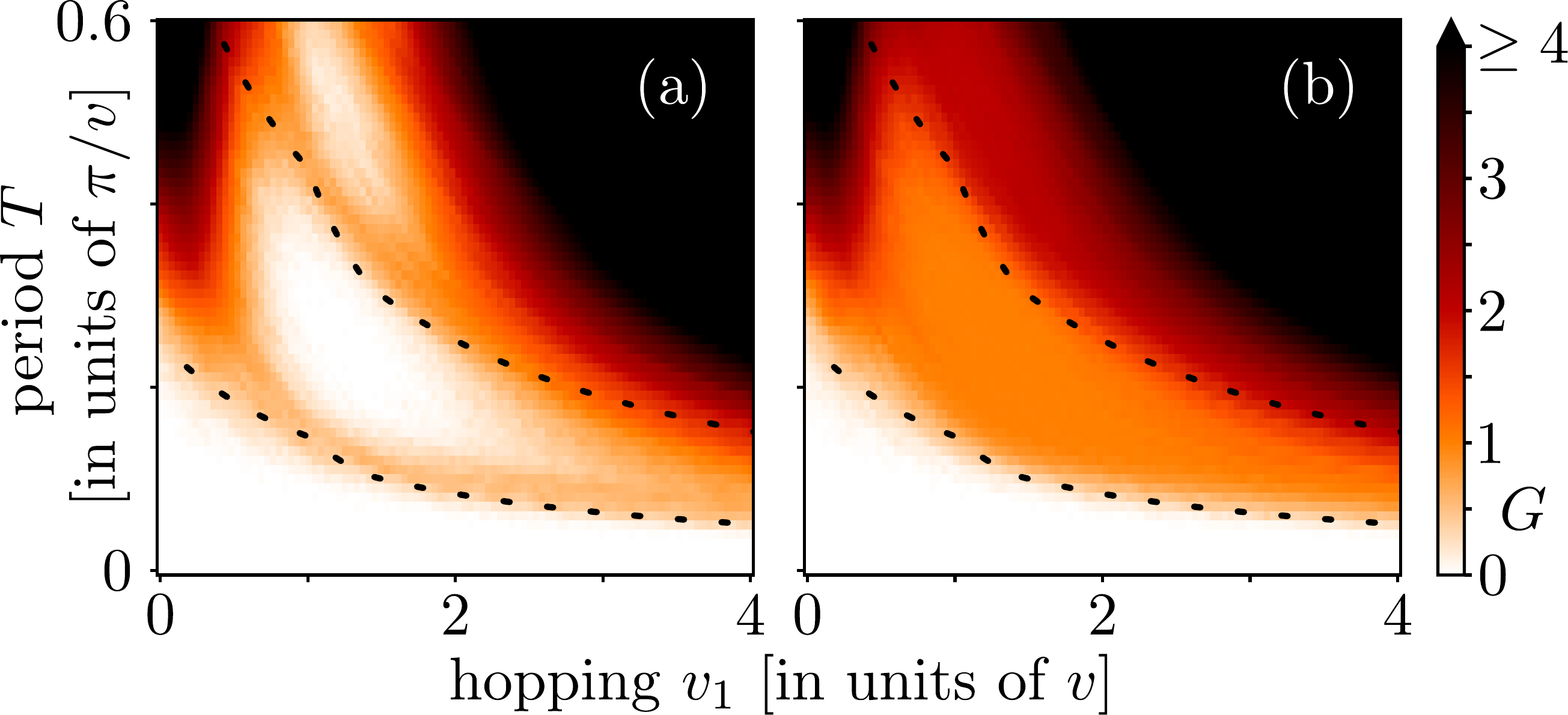}
\caption{
Phase diagram of the transmission with maximal disorder, $w=\pi$. 
With (a) periodic boundary conditions, white regions of low transmission, separated by ridges, are the AI and AFAI phases. 
The phase boundaries are well approximated by the analytical predictions (dotted lines), where $w=0$ gaps have equal sizes. 
With (b) open boundary conditions, we can read off the topological invariants of the Anderson localized phases via the quantized value of transmission.
The system size is $20\times 20$, and each point is obtained by averaging over 50 disorder realizations.
}
\label{fig:tt1}
\end{figure}

We find a simple rule of thumb to predict whether maximal disorder ($w=\pi$) leads to an AI or an AFAI: the winding number $\mathcal{W}$ of the fully localized phase at $w=\pi$ is given by the winding number of the dominant gap in the case without disorder, $w=0$.
We thus expect phase transitions between AI and AFAI to occur at $T= (2m+1) T_c$, with $m\in\mathbb{N}$ and $T_c$ given by Eq.~\eqref{eq:T_c}.
This is already seen in the data of Fig.~\ref{fig:tw}, where the dashed lines showing the expected transitions agree well with the data.
However, it also holds in the more general case, with  $v_1 \neq v$ in the Hamiltonian of Eq.~\eqref{eq:ham}, as shown in Fig.~\ref{fig:tt1}.
Here, we again find good agreement between the numerically obtained phase transitions and the condition $T= (2m+1) T_c$, now with $T_c$ depending on $v_1$ in a piecewise linear fashion (see Supplemental Material \cite{suppmat}).

\emph{Conclusion and discussion} --- 
We have shown that one can realize the AFAI, \emph{i.e.}, full Anderson localization in the bulk and topologically protected edge states, by adding disorder in the form of onsite potential kicks to a Chern insulator.
The transition to the AFAI phase takes place via an anomalous form of the levitation-and-annihilation of extended states carrying the Chern numbers, different from previously studied cases \cite{Titum2016, Rodr_guez_Mena_2019, woo2019}.
The winding number of the fully disordered $w=\pi$ system is simply given by the winding number of the largest quasienergy gap at $w=0$.
It would be interesting to explore whether this simple rule still holds in models with more than two bands.
It would also be interesting to consider this process in different symmetry classes, where weak antilocalization can lead to metallic phases, in higher-order topological insulators \cite{Benalcazar2017, Benalcazar2017b, Schindler2018, Khalaf2018, Trifunovic2019}, or in quantum walks \cite{Edge2015, Sajid2019}.
Finally, we believe that our approach of using onsite potential ``kicks'' might offer an experimentally more viable route towards AFAI phases than the ones relying on more complicated, multi-step driving protocols.

\emph{Acknowledgements} --- We thank Ulrike Nitzsche for technical assistance, and Tibor Rakovszky for discussions.
This work is supported by the Deutsche Forschungsgemeinschaft (DFG, German Research Foundation) through the W\"{u}rzburg-Dresden Cluster of Excellence on Complexity and Topology in Quantum Matter -- \emph{ct.qmat} (EXC 2147, project-id 39085490).
J.K.A. acknowledges support from the National Research, Development and Innovation Fund of Hungary within the Quantum Technology National Excellence Program (Project No.~2017-1.2.1-NKP-2017-00001), and FK 124723 and FK 132146.

\bibliography{Refs}

\begin{thebibliography}{49}
\bibitem [{par()}]{params1}%
  \BibitemOpen
  \href@noop {} {}\bibinfo {note} {{T}he transmission $G$ is obtained for
  system sizes $60\times60$ and $v_1=v$ by averaging over 1000 disorder
  realizations. {L}eft and right panels show results for $T=0.1\pi/v$ and
  $0.2\pi/v$, respectively.}\BibitemShut {Stop}%
\bibitem [{sup()}]{suppmat}%
  \BibitemOpen
  \href@noop {} {}\bibinfo {note} {See the {S}upplemental {M}aterial at [{URL}
  will be inserted by publisher] for additional details about the system{'}s
  bandstructure, localization properties, and scattering matrix
  invariants.}\BibitemShut {Stop}%
\end{thebibliography}

\end{document}


\title{Supplemental Material to ``Anomalous levitation and annihilation in Floquet topological insulators''}
\author{Hui Liu}
\affiliation{IFW Dresden and W{\"u}rzburg-Dresden Cluster of Excellence ct.qmat, Helmholtzstrasse 20, 01069 Dresden, Germany}
\author{Ion Cosma Fulga}
\affiliation{IFW Dresden and W{\"u}rzburg-Dresden Cluster of Excellence ct.qmat, Helmholtzstrasse 20, 01069 Dresden, Germany}
\author{J\'{a}nos Asb\'{o}th}
\affiliation{Institute for Solid State Physics and Optics, Wigner Research Centre for Physics, Hungarian Academy of Sciences, H-1525 Budapest P.O. Box 49, Hungary}

\begin{abstract}
In this Supplemental Material, we include some numerical and analytical results that support or illustrate how various anomalous Floquet-Anderson insulator (AFAI) phases are reached by adding disordered onsite kicks at regular time intervals to a Chern insulator. 
For a ribbon geometry with no disorder, we show how the dispersion relation and the transmission eigenvalues depend on the period time. 
We also include the analytical calculations of the minima and maxima of the energy bands. 
For the case of maximal disorder, we investigate numerically two examples of AFAI phases with winding number of $\mathcal{W}=1$ and $\mathcal{W}=2$, calculating the transmission eigenvalues and the winding of the reflection matrix.
\end{abstract}
\maketitle

\section{Floquet spectrum of the clean system with edge states}

For any Chern insulator, when observed only periodically, the mere choice of period $T$ can change the topological invariants of the Floquet spectrum, even though the Chern numbers do not depend on $T$.
We discussed this in the main text, and illustrated it in the repeated Floquet zone representation in Fig.~3.
We now complement this discussion with bandstructure plots of the Floquet model discussed in the main text, using a ribbon geometry (infinite along ${\bf a}_x$), for various values of the period $T$, in the first Floquet zone.

\begin{figure}
\centering
\includegraphics[width=1\linewidth]{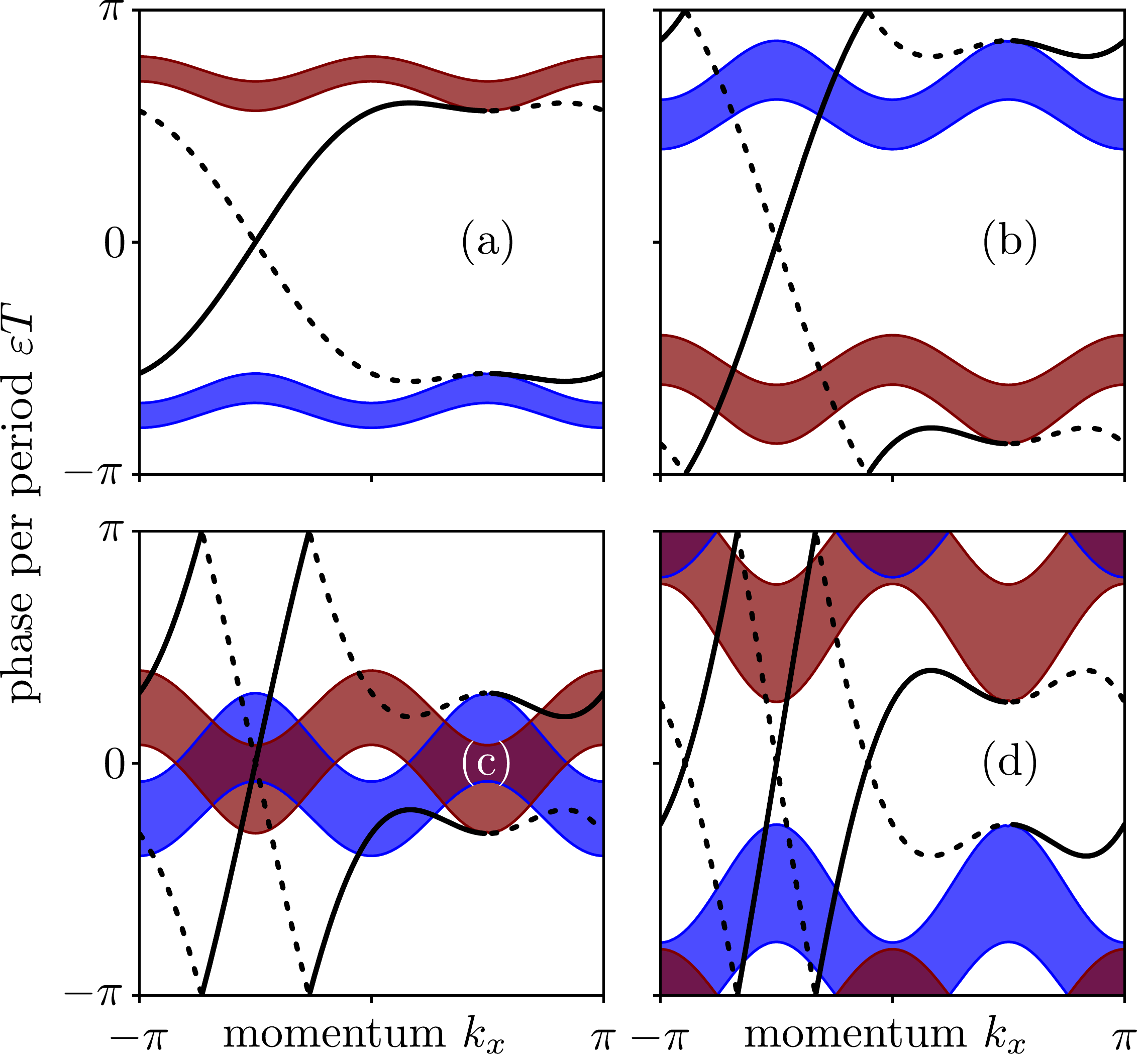}
\caption{
Quasienergy spectrum of the Floquet operator $\hat{\cal F}$ of Eq.~(4) of the main text, setting $v_1=v$ and $w=0$.
A ribbon geometry was used (infinite along ${\bf a}_x$, $60$ unit cells along ${\bf a}_y$), with bulk bands colored according to whether they correspond to upper or lower bands in the Hamiltonian $\hat{H}_0$, and edge states on the upper/lower edge shown in dashed/solid lines. 
In (a), the period is short, $T=0.2\pi/v$, and the quasienergy spectrum is the same as the spectrum of the Hamiltonian (cf.~Fig.~2 of the main text). 
In (b), with $T=0.4\pi/v$, the quasienergy bands have crossed the Floquet zone boundary. 
In (c), with $T=0.6\pi/v$, the gap around $\varepsilon=0$ is closed, while in (d), with $T=0.8\pi/v$, the gap at the Floquet zone boundary is closed.
\label{fig:sm1}
}
\end{figure}

In the clean limit, $w=0$, we show four examples of quasienergy spectra in Fig.~\ref{fig:sm1}, with different values of period time, (a): $T=0.2\pi/v$; (b): $T=0.4\pi/v$; (c): $T=0.6\pi/v$; (d): $T=0.8\pi/v$.
In case (a), the phase picked up during one period for all energy eigenstates is $\left|{T\varepsilon}\right|<\pi$.
Here the Floquet spectrum is the same as the static spectrum (Fig.~2 in the main text), with the two bulk bands and the two edge states in between.
In case (b), the period has been increased to $T=0.4 \pi/v$, the Floquet zone has shrunk, and thus both quasienergy bands have crossed the $\varepsilon=\pm \pi/T$ boundaries.
As a result, the edge states now wind in quasienergy.
In case (c), we have a still larger period, $T=0.6\pi/v$, and the quasienergy gap around $\varepsilon=0$ closes.
In case (d), with $T=0.8\pi/v$, this gap has reopened, and there are 3 edge states crossing it on both edges.

\begin{figure}
\centering
\includegraphics[width=1\linewidth]{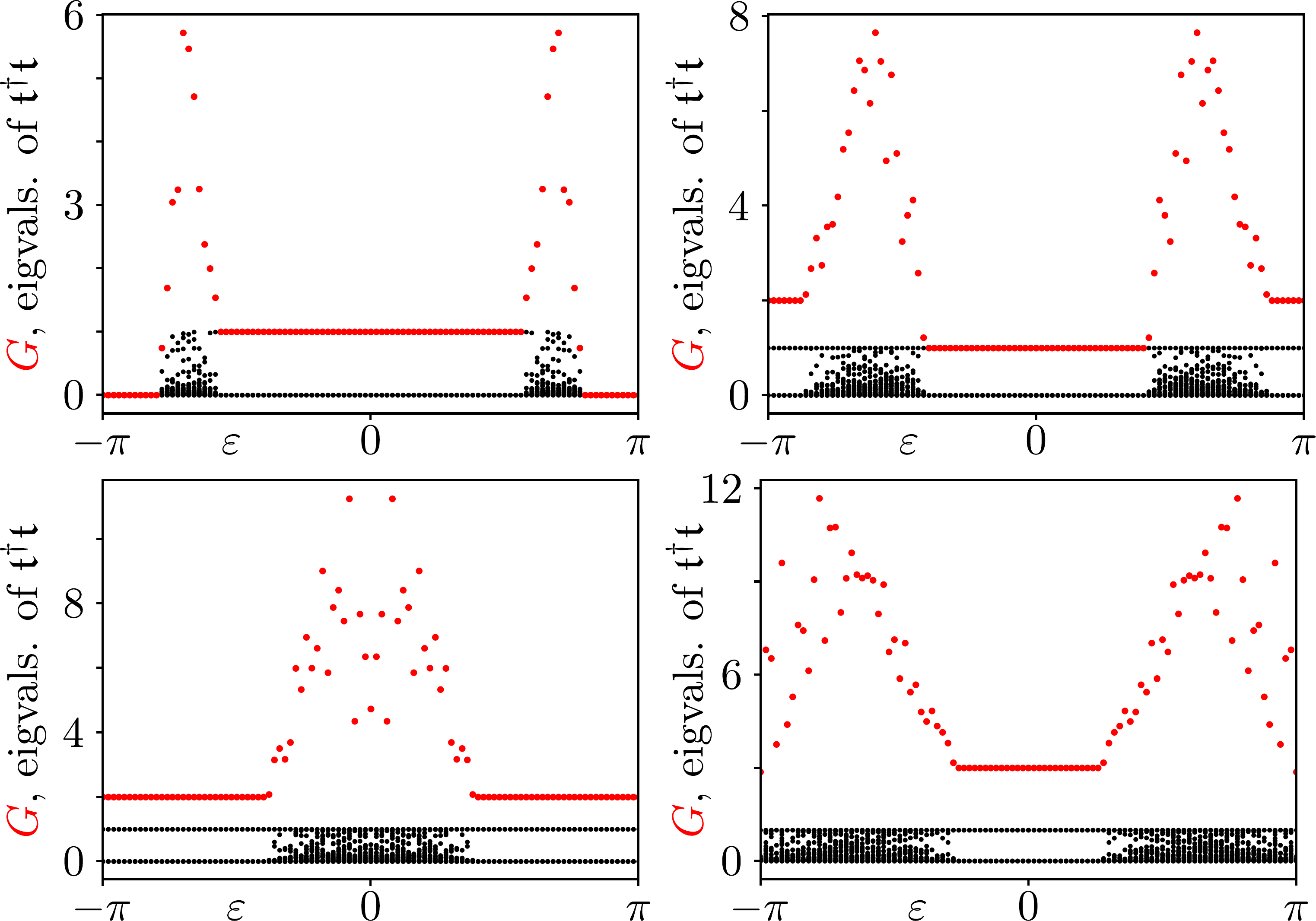}
\caption{
Total transmission $G(\varepsilon)$ (red) and individual transmission eigenvalues (black) as a function of quasienergy for the Floquet system $\hat{\cal F}$ with open boundary conditions, setting $v_1=v$ and $w=0$.
The period times are (a) $T = 0.2\pi/v$, (b) $T= 0.4\pi/v$, (c) $T=0.6\pi/v$ and (d) $T=0.8 \pi/v$.
Edge states are completely open channels with transmission eigenvalues of $1$, and bulk states are partially open channels (transmission eigenvalues $<1$) for quasienergies in bulk bands, or closed channels (vanishing transmission eigenvalues) for quasienergies in the bulk gaps. 
A system size of $20\times 20$ unit cells was used.
}
\label{fig:sm2}
\end{figure}

We show the transmission eigenvalues, \emph{i.e.}, the contributions of all channels to the transmission $G(\varepsilon)$, calculated at all quasienergies for the four cases (a-d) of the previous paragraph, in Fig. \ref{fig:sm2}.
For quasienergy values in the bulk gaps, for all the cases (a-d), the transmission eigenvalues (black), \emph{i.e.}, eigenvalues of $\mathfrak{t}^\dagger \mathfrak{t}$, are 0 or 1, corresponding to completely closed channels of transmission across the bulk, and completely open channels of transmission via edge states.
In the bulk gaps, the total transmission $G = \mathrm{tr} \left( \mathfrak{t}^\dagger \mathfrak{t} \right)$ (red) is thus just the number of completely open edge state channels.
For quasienergy values in the bulk bands, transmission eigenvalues take values between 0 and 1, corresponding to partially open channels across bulk.
The total transmission $G$ is here not quantized, and has no obvious relation to the number of channels.

\section{Band center of the model Hamiltonian with general parameters}

Our analytical predictions for the phase transitions between AI and AFAI phases in the main text relied on the band center, computed as the average of the maximum and minimum energy values of the upper band of the Hamiltonian.
In this Section we detail how we obtained these analytically.

We write the Hamiltonian of Eq.~(1) in the main text in quasimomentum-space, in a sublattice basis.
It is a $2\times 2$ matrix of the form
\begin{equation}
  \hat{H}_0 (k_x,k_y) = \pmb{h}(k_x,k_y)\cdot \hat{\pmb{\sigma}},
\end{equation}
with 
 \begin{equation}
 \begin{split}
 h_{1}(k_x,k_y)+ih_{2}(k_x,k_y)=&v_1 (1-i)[1+e^{i(k_x+k_y)}]\\&+v_1 (1+i)[e^{ik_x}+e^{ik_y}],\label{B1}\\
 h_{3}(k_x,k_y)=&2v(\cos k_x-\cos k_y).
 \end{split}
 \end{equation}
Here the $\hat{\pmb{\sigma}}$ is the formal vector composed of the $2\times 2$ Pauli matrices describing the sublattice degree of freedom.

Since the energy in the upper band, $E_+(k_x,k_y)$, is a smooth function of the quasimomenta $k_x$ and $k_y$ in the Brillouin zone, we can find its local maxima and minima from stationary points.
A stationary point of an analytical function $E_+$ is defined by
\begin{align}
  \frac{\partial E_+}{\partial k_x} &= 0; \quad\text{and}&
  \frac{\partial E_+}{\partial k_y} &=0.\label{sp}
\end{align}
A stationary point corresponds to a local extremum if the discriminant of the Hessian is positive, \emph{i.e.},
\begin{equation}
  \frac{\partial^2 E_+}{\partial k_x^2}\frac{\partial^2 E_+}{\partial k_y^2}
  - \left( \frac{\partial^2 E_+}{\partial k_x\partial k_y}\right)^2 > 0.
    \label{eq:judge}
\end{equation}
A stationary point that is a local extremum is a local minimum or maximum depending on the sign of the second derivative of the energy along any of the quasimomenta, \emph{i.e.}, maximum if
\begin{equation}
  \frac{\partial^2 E_+}{\partial k_x^2}<0,
\label{eq:max}
\end{equation}
and minimum otherwise.

In our case, the stationary points are at 8 highly symmetric points in the Brillouin zone: 4 points where $k_x$ and $k_y$ is either 0 or $\pi$, and 4 points where $k_x$ and $k_y$ are $\pm \pi/2$.
We obtain
\begin{eqnarray}
\text{max}(E_+)&=&\begin{cases}4v, \quad \text{if~}0 < v_1< v;\\
  4v_1,\quad \text{if~} v < v_1;\end{cases}\\
\text{min}(E_+)&=&\begin{cases}2 \sqrt{2} v_1,\quad \text{if~} 0 < v_1 < \sqrt{2}v;\\
  4v, \quad \text{if~} \sqrt{2} v < v_1. \end{cases}
\end{eqnarray}
We can then calculate the band center,
\begin{align}
  E_{1/2}&=\frac{\text{max}E_++\text{min}E_+}{2},
\end{align}
and find,
\begin{align}
  E_{1/2} &=\begin{cases}
  \sqrt{2} v_1 + 2v, \quad \text{if~}0 < v_1< v;\\
  (2+\sqrt{2}) v_1, \quad \text{if~} v < v_1< \sqrt{2}v;\\
  2v_1 + 2 v, \quad \text{if~} \sqrt{2}v < v_1.
\end{cases}
\end{align}

\section{Transmission and topology in the fully disordered limit}

We have shown that adding completely disordered onsite potential kicks to a Chern insulator Hamiltonian can result in AFAI phases with different topological invariants (winding numbers) $\mathcal{W}$.
We now further substantiate these findings with a more in-depth numerical study of two concrete examples of AFAI, with winding numbers of 1 and 2.
Both examples are obtained from the Hamiltonian $\hat{H}_0$ of the main text, with $v_1=v$, and with onsite potential kicks with maximal disorder $w=\pi$.
The difference between the two examples is in the time period $T$ separating the kicks:
\begin{itemize}
  \item[(a)] $T=0.29 \pi/v\simeq 2T_c$ leading to $\mathcal{W}=1$, and
  \item[(b)] $T=0.58\pi/v\simeq 4T_c$ leading to $\mathcal{W}=2$.
   \end{itemize}
We chose period times close to $T\approx 2T_c$ and $T \approx 4T_c$, where there is only one, large gap in the $w=0$ case, and therefore we expect the localization length in the fully disordered $w=\pi$ case to be short. 
We now characterize these different AFAI phases by calculating the quasienergy-dependent total transmissions, and the winding numbers of the reflection matrix.

\begin{figure}
\centering
\includegraphics[width=1\linewidth]{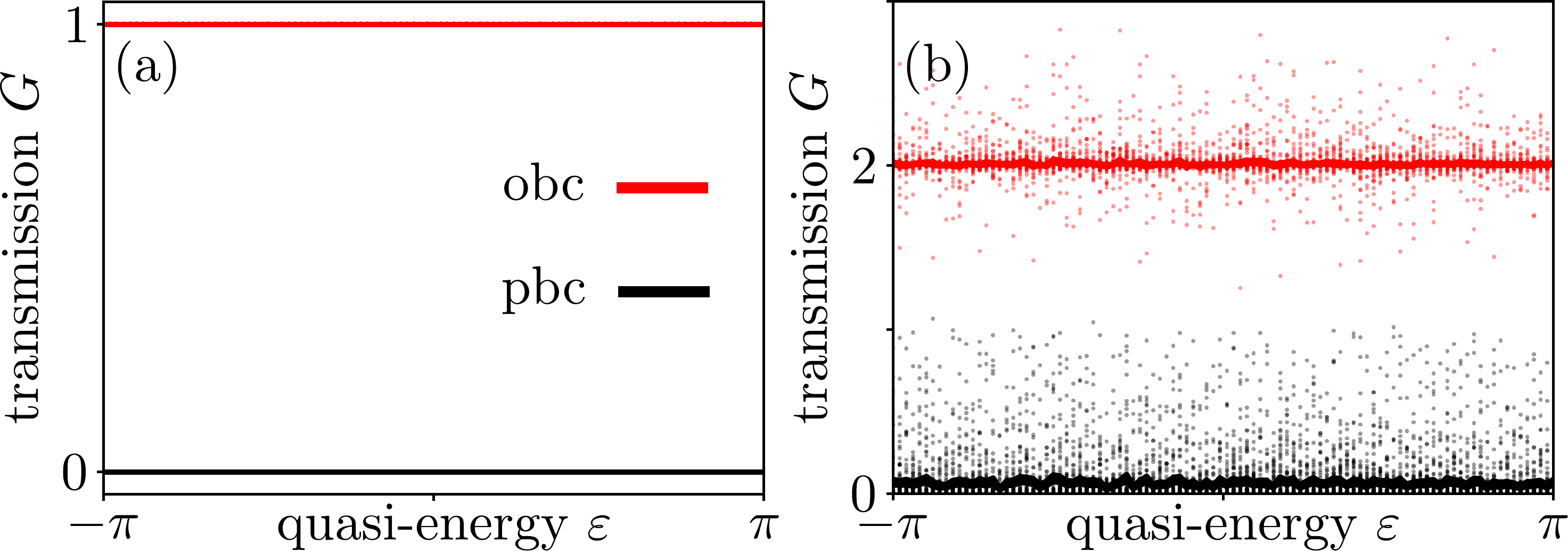}
\caption{
Quasienergy-dependent transmission $G(\varepsilon)$ of the periodically kicked system, with $v_1=v$ and $w=\pi$, setting (a) $T=0.29\pi/v$, and (b) $T=0.58\pi/v$. 
The transmission was calculated with both periodic and open boundary conditions, for 100 disorder realizations on systems of size $100\times 100$ unit cells. 
The disorder-averaged transmission (continuous lines, thickness represents error bars) indicates Anderson localization of the bulk and topological invariants of 1 in (a) and 2 in (b), at all quasienergies.
For the case (b), we observe relatively large sample-to-sample fluctuations of the transmission values both with open (red dots) and with periodic boundary conditions (black dots).
}
\label{fig:afai}
\end{figure}

We begin with the total transmissions, $G(\varepsilon)= \mathrm{tr} \left[\,\mathfrak{t}(\varepsilon)^\dagger \mathfrak{t}(\varepsilon) \right]$.
We calculated these for the two cases (a) and (b) with $w=\pi$, averaging over 100 random disorder realizations for each point, on systems of size $L=100$, and show the results in Fig. \ref{fig:afai}.
The numerics indicates total Anderson localization of the bulk in both cases, since with periodic boundary conditions, the disorder-averaged transmission at all quasienergies is small in both (a) and (b).
We do see finite-size effects for case (b), with total transmission almost reaching 1 at some quasienergies with some random realizations of disorder.
Switching to open boundary conditions, we obtain a quantized increase in the disorder-averaged total transmission at every quasienergy for both cases, by 1 and by 2, corresponding to the presence of 1 or 2 chiral edge modes.
Note again the sample-to-sample fluctuations in case (b), which we attribute to finite-size effects.

\begin{figure}
\centering
\includegraphics[width=1\linewidth]{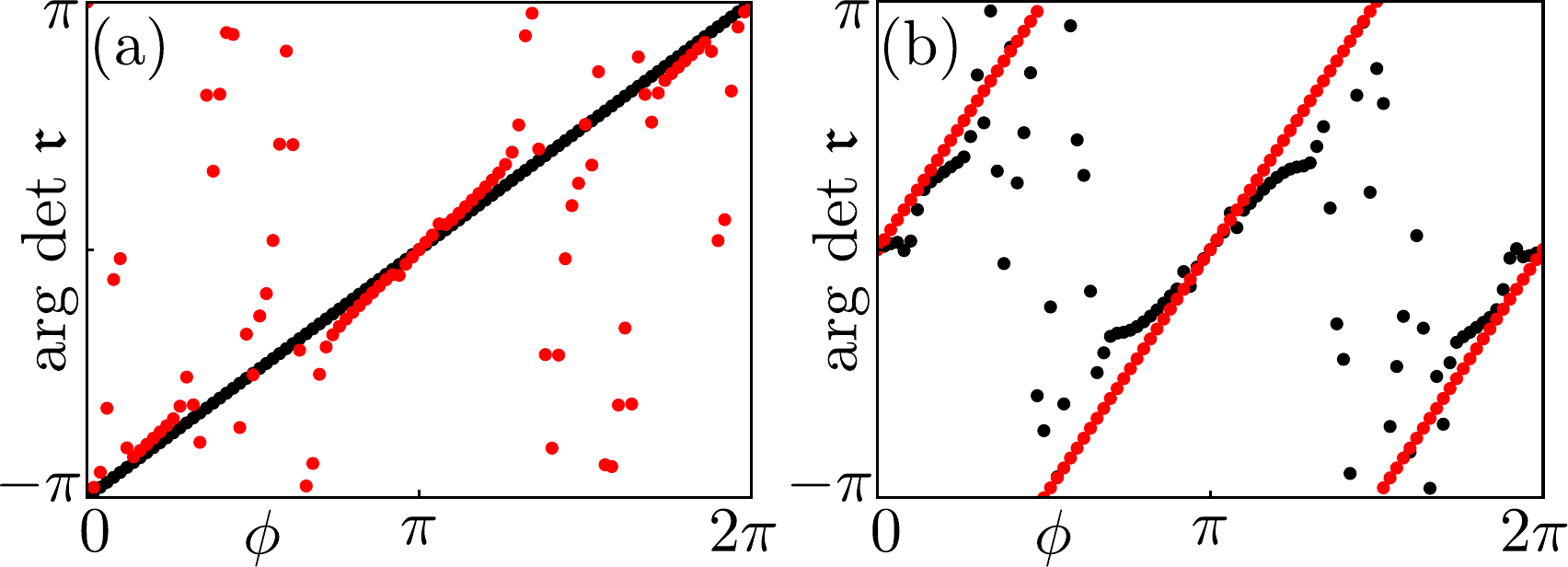}
\caption{
Clean case: Dependence of the phase of the determinant of the reflection matrix $\mathfrak{r}(\varepsilon, \phi)$ on the phase $\phi$ used to twist the boundary conditions, for the Floquet model with $v_1=v$ and  $w=0$. 
We calculated the reflections at quasienergy $\varepsilon=0$ (black) and $\varepsilon=\pi/T$ (red), for a system size of $40 \times 40$. 
In (a), $T=0.29\pi/v$, there is a gap at $\varepsilon=0$, with continuously winding value of $\det\mathfrak{r}$, but no gap at $\varepsilon=\pi/T$, and hence a discontinuous phase of $\det\mathfrak{r}$. 
In (b), $T=0.58\pi/v$, the situation is reversed, with clear winding of 2 at $\varepsilon=\pi/T$, but a discontinuous curve at $\varepsilon=0$.
}
\label{varphi_w0}
\end{figure}

\begin{figure}
\centering
\includegraphics[width=1\linewidth]{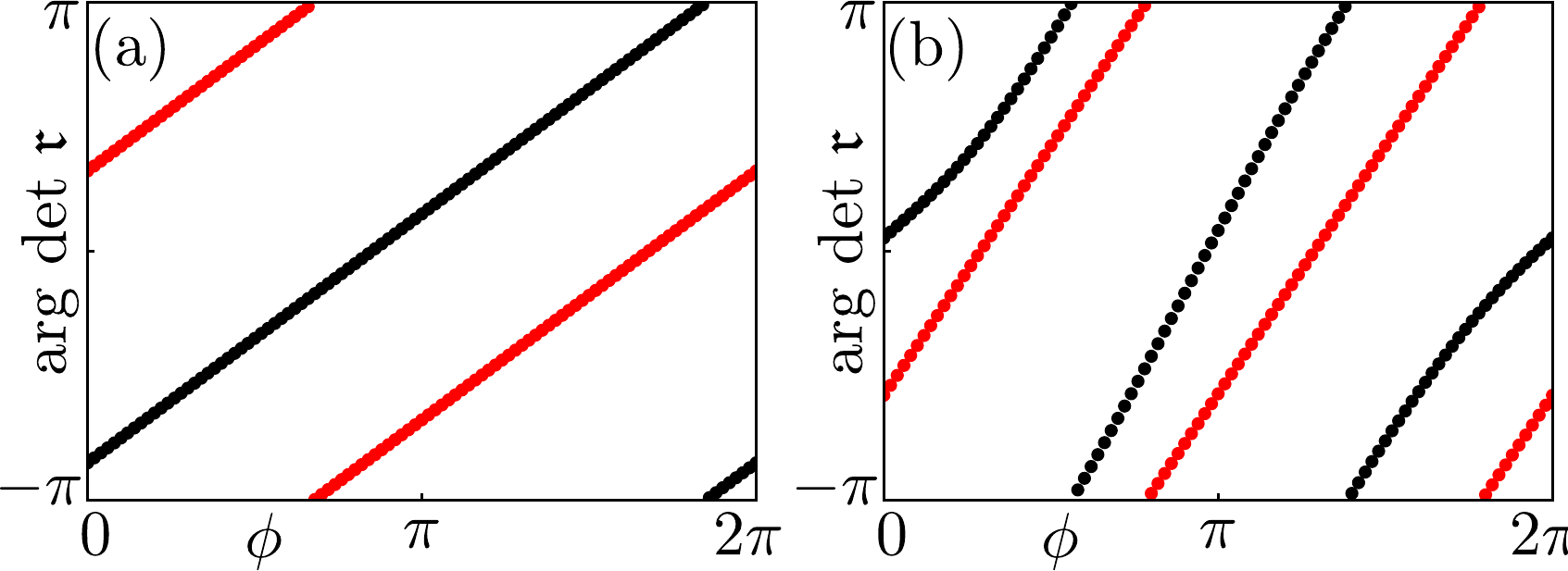}
\caption{
Disordered case: Dependence of the phase of the determinant of the reflection matrix $\mathfrak{r}(\varepsilon, \phi)$ on the phase $\phi$ used to twist the boundary conditions, for the Floquet model with $v_1=v$ and $w=\pi$. 
We calculated the reflections at quasienergy $\varepsilon=0$ (black) and $\varepsilon=\pi/T$ (red), for a single realization of disorder, using a system size of $40\times 40$. 
Because of the fully Anderson localized bulk, $\det\mathfrak{r}(\varepsilon,\phi)=1$ at all quasienergies, and hence we unambiguously observe the winding numbers: in (a), with $T=0.29\pi/v$, of $\mathcal{W}=1$; and in (b), with $T=0.58\pi/v$, of $\mathcal{W}=2$.
}
\label{varphi_wpi}
\end{figure}

\emph{Topological invariant} --- 
A robust numerical tool to compute the topological invariant of an insulating bulk (translation invariant or not) is the winding number of the reflection matrix \cite{fulga2012scattering, Fulga2016}.
The reflection matrix is the $\mathfrak{r}$ component of the scattering matrix defined in the main text.
We need to take twisted boundary conditions, \emph{i.e.}, periodic boundary conditions (cylinder geometry), but with all hopping amplitudes crossing the boundary rightward or leftward multiplied by $e^{i\phi}$ or $e^{-i\phi}$, respectively. 
This is equivalent to threading a flux $\phi$ through the cylinder.
The winding number $\mathcal{W}$ is
\begin{eqnarray}
  \mathcal{W}(\varepsilon)=\frac{1}{2\pi i}\oint_{-\pi}^{\pi}d\phi
  \frac{d}{d\phi}\text{log}~\text{det}[\mathfrak{r}(\varepsilon,\phi)].
\label{eq:ti}
\end{eqnarray} 
For $\varepsilon$ in a quasienergy gap of a clean system, this winding number gives the topological invariant of the gap \cite{Rudner2013, Fulga2016, Fulga2019}.
In the periodically kicked Hamiltonian with completely disordered kicks, $w=\pi$, if the bulk is Anderson localized, the winding number of the reflection matrix is quasienergy-independent and gives the topological invariant.

We illustrate how the calculation of the winding number works by showing how the determinant of the reflection matrix winds as the phase $\phi$ is tuned, for the cases (a) and (b) above.
We first show this winding for the clean case, $w=0$, at quasienergies in the middle and at the boundary of the Floquet zone, $\varepsilon=0$ and $\varepsilon=\pi/T$, in Fig.~\ref{varphi_w0}. 
In both cases, there is a gap at only one of these quasienergies, and not at the other one. 
Thus, the phase of $\det \mathfrak{r}$ is continuous in the case (a) at $\varepsilon=0$, but not at $\varepsilon=\pi/T$, and the reverse is true in the case (b). 
The winding numbers of the continuous $\det \mathfrak{r}$ curves are easily read off, and give the expected values of 1 and 2.
We show the same quantities for the case with complete disorder, $w=\pi$, in Fig.~\ref{varphi_wpi}. 
Due to complete Anderson localization, both cases have bulk mobility gaps, and smooth curves of ${\rm arg} \det \mathfrak{r}$ at all quasienergies. 
The winding numbers of the curves are $\mathcal{W}=1$ and $\mathcal{W}=2$, in line with our expectations. 

\bibliography{Refs}